\documentclass[twocolumn,prl,aps,preprint]{revtex4}
\newcommand{\be}{\begin{equation}} \newcommand{\ee}{\end{equation}} 
\newcommand{\bea}{\begin{eqnarray}}\newcommand{\eea}{\end{eqnarray}}

\begin{document}
\preprint{SINP/TNP/04-16, cond-mat/0410398}
\title{ Comment on ``Dynamical role of anyonic excitation statistics in
rapidly rotating Bose gases"}
\author{ Pijush K. Ghosh}
\email{pijush@theory.saha.ernet.in}
\affiliation{Theory Division, Saha Institute of Nuclear Physics, 
Kolkata 700 064, India.}
\begin{abstract} 
We comment on the work of Fischer, Phys. Rev. Lett. 93, 160403 (2004).
Contrary to the claim in the letter, we argue that the anyonic
excitation statistics does not play any dynamical role in
stabilizing attractive bose gases under rapid rotation in
a fractional quantum Hall state. We also point out that the
assertion of obtaining exact solutions of the self-dual equation
that saturates the Bogomol'nyi bound is invalid for non-zero
external field. 
\end{abstract}
\maketitle

It has been claimed in a recent letter that attractive bose gases under
rapid rotation can be stabilized in a fractional quantum Hall state
due to the anyonic statistics of their quasiparticle excitations\cite{fis}.
Further, it is claimed that the solutions of the self-dual equation
satisfying Bogomol'nyi bound with non-zero ``external" field can be obtained
exactly from that of Jackiw-Pi solutions through time-dependent coordinate
transformations\cite{fis}. We argue in this comment with the help of known
results\cite{pij1,pij2} that the anyonic excitation statistics does not play
any dynamical role in stabilizing the attractive bose gases. We also point out
that the assertion of obtaining exact solutions of the self-dual equation (11)
in Ref. \cite{fis} with $\Omega \neq 0$ is invalid.

The dynamical stability of the system described by the Hamiltonian $H$ in
Ref. \cite{fis} with quartic self-interaction has been studied
previously\cite{pij1}. Results obtained in subsequent papers\cite{pij2,pij3}
are also equally valid for $H$ due to an underlying universality. The
mean-square radius $I=\frac{m}{2} \int d^2 r^2 \psi^* \psi$ has the same
dynamical behaviour for the system with or without the Chern-Simons
gauge field, if it is evolved with the same set of initial conditions and a
fixed $H > 0$ for both the cases\cite{pij1,pij2}. On the other hand, the
field $\psi$ blows up (i.e. $I$ collapses) at a finite time independent of
initial conditions if $H \leq 0$\cite{pij1,pij2}. This result is again valid
for both in presence and in absence of the Chern-Simons gauge field.

Analyzing Eq. (8) of
Ref. \cite{fis} with an attractive quartic self-interaction( i.e.
${\cal{U}}=  g {\mid \psi \mid}^4$ with $g < 0$), at the self-dual point
$g_{eff}=0$ and $V(x)=0$, we find that $H$ is
negative for $\Theta \Omega > 0$. The fractional quantum Hall
state with the filling factor $\nu < 1$ is obtained in Ref. \cite{fis} for
both $\Theta$ and $ \Omega$ positive, implying $H < 0$. Hence, the system is
not dynamically stable and the field $\psi$ blows up at a finite time.
In fact, there are no finite energy solutions of the self-dual Eqn. (11) for
$ g < 0$ and $\Theta \Omega > 0$, since the energy-functional is not bounded
from below.

Our second point concerns about the claim of obtaining exact solutions of
the self-dual equation (11) with $\Omega \neq 0$.
Any solution of a self-dual equation saturating the Bogomol'nyi bound is
necessarily a solution of the second order equations of motion that are obtained
by varying the action. However, the converse is not true. There are solutions
of the second order equations of motion with higher energy which are not
solutions of the self-dual equation. The mapping that relates $S_{\Theta}$ with
$\Omega \neq 0$ to that $S_{\Theta}$ with $\Omega=0$ is only at the level of
action. Thus, an exact solution of the second order equations of motion of
$S_{\Theta}$ at $g_{eff}=0$, can definitely be obtained from Jackiw-Pi soliton
of $S_{\Theta}$ with $\Omega=0$, which however is not a solution of the
Eqn. (11). In fact, unlike these exact solutions, the density
$\rho=\psi^{*} \psi$ of the self-dual system is not determined by the
Liouville equation for $\Omega \neq 0$\cite{ezawa}.

This work is supported by SERC, DST, Govt. of India.


\begin{thebibliography}{99}
\bibitem{fis} U. R. Fischer, Phys. Rev. Lett. {\bf 93}, 160403 (2004).
\bibitem{pij1} Pijush K. Ghosh, Phys. Rev. {\bf A65}, 012103 (2002).
\bibitem{pij2} Pijush K. Ghosh, Phys. Rev. {\bf A65}, 053601 (2002).
\bibitem{pij3} Pijush K. Ghosh, Phys. Lett. {\bf A308}, 411 (2003).
\bibitem{ezawa} Z. F. Ezawa, M. Hotta and A. Iwazaki, 
Phys. Rev. {\bf D44}, 452 (1991).
\end{thebibliography}
\end{document}